%% file: main.tex
\documentclass[sigconf]{acmart}

\include{dlab}
\usepackage{subcaption}
\usepackage{hyperref}
\usepackage{balance}
\usepackage{footmisc}

\AtBeginDocument{%
  \providecommand\BibTeX{{%
    \normalfont B\kern-0.5em{\scshape i\kern-0.25em b}\kern-0.8em\TeX}}}

\setcopyright{none}
\renewcommand\footnotetextcopyrightpermission[1]{} 
\begin{document}

\title[Going Down the Rabbit Hole: Characterizing the Long Tail of Wikipedia Reading Sessions]{Going Down the Rabbit Hole: \\Characterizing the Long Tail of Wikipedia Reading Sessions}

\author{Tiziano Piccardi}
\affiliation{%
  \institution{EPFL}
  \country{}
  }
\email{tiziano.piccardi@epfl.ch}

\author{Martin Gerlach}
\affiliation{%
  \institution{Wikimedia Foundation}
  \country{}
  }
\email{mgerlach@wikimedia.org}

\author{Robert West}
\affiliation{%
  \institution{EPFL}
  \country{}
  }
\email{robert.west@epfl.ch}



\begin{abstract}
``Wiki rabbit holes'' are informally defined as navigation paths followed by Wikipedia readers that lead them to long explorations, sometimes involving unexpected articles. Although wiki rabbit holes are a popular concept in Internet culture, our current understanding of their dynamics is based on anecdotal reports only. To bridge this gap, this paper provides a large-scale quantitative characterization of the navigation traces of readers who fell into a wiki rabbit hole. First, we represent user sessions as navigation trees and operationalize the concept of wiki rabbit holes based on the depth of these trees. Then, we characterize rabbit hole sessions in terms of structural patterns, time properties, and topical exploration.
We find that article layout influences the structure of rabbit hole sessions and that the fraction of rabbit hole sessions is higher during the night. Moreover, readers are more likely to fall into a rabbit hole starting from articles about entertainment, sports, politics, and history. Finally, we observe that, on average, readers tend to stay focused on one topic by remaining in the semantic neighborhood of the first articles even during rabbit hole sessions.
These findings contribute to our understanding of Wikipedia readers' information needs and user behavior on the Web.

\end{abstract}



\maketitle

\section{Introduction}
If you ever visited Wikipedia for simple fact-checking and then, in a few minutes, found yourself learning that \textit{life does not evolve wheels}\footnote{\url{https://en.wikipedia.org/wiki/Rotating_locomotion_in_living_systems}} and that \textit{unsold video games may end up buried in the desert}\footnote{\url{https://en.wikipedia.org/wiki/Atari_video_game_burial}}, there is a high chance that you fell into a ``wiki rabbit hole'', a term inspired by Lewis Carroll's novel \textit{Alice's Adventures in Wonderland,} where the main character reaches an astonishing world by following a white rabbit deep into its burrow.


Similarly, a wiki rabbit hole, sometimes called a wiki hole or wiki black hole\footnote{\url{https://en.wikipedia.org/wiki/Wiki_rabbit_hole}}, is a popular concept in Internet culture often described as a long navigation session where readers, following multiple links, get lost in Wikipedia and learn about a diverse set of topics. 
The reason that motivates readers to engage in long navigation sessions, with jumps across different topics, is often associated with boredom \cite{singer_why_2017} and a busybody-style of curiosity~\cite{Lydon-Staley2021}.

Given the substantial time we spend consuming online content,  understanding the dynamics of how we seek knowledge can offer useful insight into our information needs and support the design of better systems centered around users' interests.
Previous work provided an overall characterization of reading sessions on Wikipedia \cite{piccardi2021large}. However, reading sessions are typically short (78\% consisting of a single pageload) such that the population-wide average does not adequately capture the behavior contained in the long tail. Therefore, this study serves as a complimentary analysis of how readers browse Wikipedia by focusing on long reading sessions associated with rabbit hole navigation.


This paper characterizes the long tail of these navigation traces by focusing on sessions with paths generated by at least ten sequential internal clicks.
We use server logs collected for one month from English Wikipedia to describe the exploration dynamics and how readers navigate during the long reading sessions. 
We characterize the paths of readers who engage in extended navigation by describing common \textit{structural properties} such as the shape of the navigation traces, \textit{temporal patterns} such as the time spent on Wikipedia and the daily rhythm, and \textit{topic-based characteristics} like the semantic diffusion of the navigation from the origin.

This paper is organized as follows. First, we introduce the related work (\Secref{sec:related_work}) and the data (\Secref{sec:data}). Then, we operationalize wiki rabbit holes and  describe the characteristics of these sessions in terms of structure (\Secref{subsec:structural}), time properties (\Secref{subsec:temporal}), and topics (\Secref{subsec:topical}). Finally (\Secref{sec:discussion}), we summarise the findings and discuss the implications.

\section{Related Work}
\label{sec:related_work}
In recent years, readers' behavior on Wikipedia has been explored in different context by characterising interaction with citations \cite{piccardi2020quantifying, maggio2020meta}, external links \cite{piccardi2021gateway}, images \cite{rama2022large}, and navigation behaviors.

Readers' navigation from article to article received significant attention from researchers using different approaches such as wikigames \cite{Wikispeedia}, browser history sharing \cite{Lydon-Staley2021}, server logs, and public clickstream \cite{Wulczyn2015clickstream,arora2022wikipedia,rodi2017search}. These analyses address two different types of navigation on Wikipedia: natural and targeted navigation. 

Natural navigation refers to the typical \textit{out-of-lab} usage of Wikipedia; the digital traces left by the readers are associated with self-motivated learning or to satisfy personal information needs \cite{singer_why_2017}.
Previous studies \cite{piccardi2021large,SearchNavigationWikipedia} shows that readers in a natural setup tend to have short sessions, with the vast majority (78\%) of the sessions composed by a single pageload. When the navigation goes beyond the first page, an analysis based on the server logs revealed the users' propensity to stop the exploration when they reach a low-quality article. In other investigations, researchers found that Wikipedia articles relay different traffic volumes based on their topics \cite{dimitrov2019different} and the type of page \cite{Gildersleve2018Inspiration}. Readers also show preferences for links that appear at the top of the page and are semantically closer to the current article \cite{LinkSuccessfulWikipedia,StructureArticlesNavigation}. Reading preferences were shown to fall into 4 types of behaviors described as focus, trending, exploration and passing\cite{lehmann2014reader}.

In contrast, targeted navigation typically involves instructions for the reader to reach a specific article on Wikipedia. For example, wikigames \cite{Wikispeedia} lets users navigate on a gamified platform to reach a target article in as few clicks as possible starting from a random Wikipedia page. This line of research aims to understand how humans navigate information networks and what strategies they employ to find a predetermined piece of information. Researches found \cite{HumanWayfinding,Helic} that initially, users tend to jump to high degree nodes that act as navigational hubs, and then they converge to the destination in increasingly small steps in the semantic space. Given the clear definition of success in a wikigame --reaching the target article--, researchers also explored what makes people leave the navigation, discovering that diverging in topic space from the target leads to frustration and giving up.

Finally, a study based on a combination of voluntary sharing of the navigation history and survey-based qualitative methods \cite{Lydon-Staley2021} analyzed curiosity as a driving mechanism for navigation. They found that different curiosity patterns lead to distinct navigation behaviors when looking at the knowledge networks constructed by the readers.

\section{Data}
\label{sec:data}
To characterize the sessions of readers that fall into a wiki rabbit hole, we rely on server logs collected over four weeks in March 2021. The data is pre-processed as described in previous work \cite{piccardi2021large}, e.g., we removed bots and all the activities of users that were logged in, edited at least one article during the data collection, or accessed Wikipedia from countries with less than 300 pageloads per day in order to preserve readers' privacy. Then, we made the requests from different countries comparable by converting the timestamps in local time and dropped all the sensitive information such as geo-location, user-agent, and IP addresses.
To combine requests coming from the same client, we preserve for each request an anonymous user identifier generated from the original user-agent string and the IP address. Since large organizations may have many clients sharing these two properties, some identifiers may actually represent requests coming from different readers. To mitigate this issue, we removed all the activities associated with user identifiers with more than 2800 pageloads in 4 weeks (i.e. more than 100 pageviews per day on average). The resulting dataset comprises 6.25B pageviews generated by 1.47B different user identifiers.

\xhdr{Aggregating the sessions}
We aggregate the session into navigation trees as described in previous works \cite{ParanjapeImproving,piccardi2021large}. Given the complex navigation patterns of Web users, comprised of multi-tab and backtracking behavior, the structure of the navigation path is typically a tree. To reconstruct sequences of pageviews from individual clicks, we use the HTTP referrer field that allows the browser to specify the origin of each request. First, we use this information to generate \textit{root-only} trees for all the requests coming from URLs that are not Wikipedia articles. Then, for all the loads that originated from internal navigation, we assign each request as a child of the node representing the most recent time the user loaded the source article. The resulting dataset is composed of 3.7B navigation trees.

\begin{table*}[]
\tiny
\centering
\begin{tabular}{l|lllll}
\hline
\textbf{Tree} & \begin{tabular}[t]{@{}l@{}}
 \hspace{0.0mm}Winston Churchill\\ \hspace{0.5mm}Anthony Eden\\ \hspace{1.0mm}Harold Macmillan\\ \hspace{1.5mm}Alec Douglas-Home\\ \hspace{2.0mm}Harold Wilson\\ \hspace{2.5mm}James Callaghan\\ \hspace{3.0mm}Margaret Thatcher\\ \hspace{3.5mm}John Major\\ \hspace{4.0mm}Tony Blair\\ \hspace{4.5mm}Gordon Brown\\ \hspace{5.0mm}David Cameron\\ \hspace{5.5mm}Theresa May\\ \hspace{6.0mm}Boris Johnson\\
\end{tabular} & 
\begin{tabular}[t]{@{}l@{}}
 \hspace{0.0mm}UFC 260\\ \hspace{0.5mm}UFC on ABC: Till vs. Vettori\\ \hspace{1.0mm}UFC on ESPN: Whittaker vs. Gastelum\\ \hspace{1.5mm}UFC 261\\ \hspace{2.0mm}UFC on ESPN: Reyes vs. Procházka\\ \hspace{2.5mm}UFC on ESPN: Sandhagen vs. Dillashaw\\ \hspace{3.0mm}UFC 262\\ \hspace{3.5mm}UFC Fight Night: Font vs. Garbrandt\\ \hspace{4.0mm}UFC Fight Night 189\\ \hspace{4.5mm}UFC 263\\ \hspace{5.0mm}UFC Fight Night 190\\ \hspace{5.5mm}UFC Fight Night 191\\ \hspace{6.0mm}UFC 264\\ \hspace{6.5mm}UFC Fight Night 192\\
\end{tabular} & 
\begin{tabular}[t]{@{}l@{}}
 \hspace{0.0mm}WandaVision\\ \hspace{0.5mm}Filmed Before a Live Studio Audience\\ \hspace{1.0mm}Don't Touch That Dial\\ \hspace{1.5mm}Now in Color\\ \hspace{2.0mm}We Interrupt This Program\\ \hspace{2.5mm}On a Very Special Episode...\\ \hspace{3.0mm}All-New Halloween Spooktacular!\\ \hspace{3.5mm}Breaking the Fourth Wall (WandaVision)\\ \hspace{4.0mm}Previously On\\ \hspace{4.5mm}The Series Finale\\
\end{tabular} & 
\begin{tabular}[t]{@{}l@{}}
 \hspace{0.0mm}RuPaul's Drag Race (season 1)\\ \hspace{0.5mm}RuPaul's Drag Race (season 2)\\ \hspace{1.0mm}RuPaul's Drag Race (season 3)\\ \hspace{1.5mm}RuPaul's Drag Race (season 4)\\ \hspace{2.0mm}RuPaul's Drag Race (season 5)\\ \hspace{2.5mm}RuPaul's Drag Race (season 6)\\ \hspace{3.0mm}RuPaul's Drag Race (season 7)\\ \hspace{3.5mm}RuPaul's Drag Race (season 8)\\ \hspace{4.0mm}RuPaul's Drag Race (season 9)\\ \hspace{4.5mm}RuPaul's Drag Race (season 10)\\ \hspace{5.0mm}RuPaul's Drag Race (season 11)\\ \hspace{5.5mm}RuPaul's Drag Race (season 12)\\ \hspace{6.0mm}RuPaul's Drag Race (season 13)\\
\end{tabular} & 
\begin{tabular}[t]{@{}l@{}}
 \hspace{0.0mm}Elizabeth II\\ \hspace{0.5mm}George VI\\ \hspace{1.0mm}Edward VIII\\ \hspace{1.5mm}George V\\ \hspace{2.0mm}Edward VII\\ \hspace{2.5mm}Queen Victoria\\ \hspace{3.0mm}William IV\\ \hspace{3.5mm}George IV\\ \hspace{4.0mm}George III\\ \hspace{4.5mm}George II of Great Britain\\ \hspace{5.0mm}George I of Great Britain\\ \hspace{5.5mm}Anne, Queen of Great Britain\\
\end{tabular} \\
\hline
\textbf{Count} & \multicolumn{1}{c}{850} & \multicolumn{1}{c}{555} & \multicolumn{1}{c}{503} & \multicolumn{1}{c}{452} & \multicolumn{1}{c}{406}

\end{tabular}
\caption{
The five most frequent trees result from using navigational links in the infobox and have a chain-like structure.
}
\label{table:most_frequent_paths}
\end{table*}

\section{The Wiki Rabbit Hole}
\label{sec:wikirabbithole}

\xhdr{Wiki rabbit hole operationalization} In this work, we consider as \textit{wiki rabbit hole} sessions with a navigation tree with a minimum depth of ten steps. This constraint means we include all the trees where the longest path root-leaf is at least nine sequential clicks. 
This definition ensures that large shallow trees where the readers remained around the starting article opening many tabs, in the present work, are not considered rabbit hole sessions. Given the long-tailed nature of the tree size distribution, this filter leaves us with 216M pageviews aggregated in 8.97M trees---0.24\% of the original navigation sessions. 
In comparison, this number is much larger than the roughly 40k active editors per month in English Wikipedia\footnote{\url{https://stats.wikimedia.org/\#/en.wikipedia.org/contributing/active-editors/normal|line|2-year|(page_type)~content*non-content|monthly}}, suggesting that the majority of these trees comes from actual readers.

\xhdr{Frequent entry points}
In total, 846K articles acted, at least once, as the entry point for a rabbit hole session, i.e. they appear as the root of a tree. The most frequent articles that served as starting points for long sessions are popular pages such as \textsc{Elizabeth II} (37.8K), \textsc{Deaths in 2021} (23.7K), \textsc{2020 United States presidential election} (17.6K), \textsc{Wikipedia} (15.8K), and \textsc{Prince Philip, Duke of Edinburgh} (15.0K). These pages are overall very popular and associated with events that received significant attention at the time of the data collection. In total, 11 out of the top 20 articles have some connection with the British Royal family that, in March 2021, received a burst in media attention. Interestingly, the list of the most common pages where rabbit hole sessions start includes the article \textsc{Wikipedia}. A substantial portion of readers use this page as the starting point for navigation, from which they  move to the desired destination using the internal search. By inspecting the articles loaded from \textsc{Wikipedia}, it emerges that readers continue the navigation using the blue links available in the body of the page only in 9.3\% of the cases.
\vfill

These pages are observed frequently in this list because of the attention received overall. We observe a different pattern when we normalize the number of times an article acts as the entry point for a rabbit hole by its global popularity. By limiting to pages that received at least 100 pageviews, we find \textsc{1980 Arkansas gubernatorial election} (34\% as rabbit hole entry), \textsc{It Is the Law} (33\%), \textsc{Alexander, Duke of Schleswig-Holstein-Sonderburg} (30\%), \textsc{Solidarity Party of Afghanistan} (29\%), and \textsc{2006 Florida gubernatorial election} (29\%). 

\vfill
\xhdr{Infobox navigational links} A manual inspection of the top 1000 articles that serve most frequently as an entry point for the rabbit hole reveals that many articles refer to recurrent events with seasonal repetitions, such as elections (\ie \textsc{1946 Dutch general election}), sports events (\ie \textsc{1979 NBA All-Star Game}), and award ceremonies (\ie \textsc{2019 Academy Awards}). In total, 68.4\% of the articles in this list contain the word ``election'' in their title, and 87.3\% contain a four digits year. When considering the entire dataset, the pages about elections are 0.5\% of all the entry-points articles and cover 3.4\% of the trees.

\begin{figure}[t]
    \centering
    \includegraphics[width=0.80\linewidth]{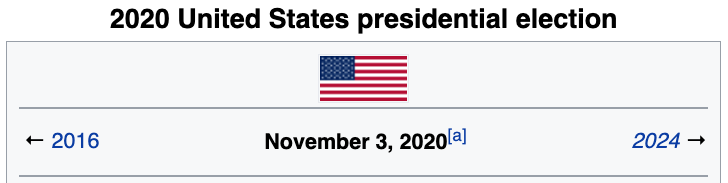}
    \caption{Navigational links in the infobox are often used to engage in long sessions.}
    \label{fig:navigation_links}
    \vspace{-3mm}
\end{figure}

By inspecting these navigation traces, we observe that readers often engage in long sessions using the navigational links available in the infobox. Articles about recurrent events typically have links to move to the previous or following occurrence of the same event. For example, a reader that follows this pattern may open the article \textsc{2020 United States presidential election} and then navigate to \textsc{1960 United States presidential election} by repeatedly clicking the year of the previous election. \Figref{fig:navigation_links} shows an example of links used for this type of navigation.

\vfill

\xhdr{Frequent pathways}
English Wikipedia in March 2021 had around 252M links between its articles\footnote{When considering the full HTML instead of the wikitext of the article this number rises to more than 400M \cite{mitrevski2020wikihist}.}. Given the large exploration space that the readers can access, it is rare that two sessions generate the same navigation tree. We observe that for the sessions with the rabbit hole pattern, 0.74\% of the trees appear more than once, and 0.03\% more than ten times. The usage of the navigational links in the infobox typically shapes these navigation traces. Table \ref{table:most_frequent_paths} shows the five most frequent paths with the respective number of occurrences. Given the readers' behavior that generated these trees, they exhibit a chain-like structure without branching. 

\vfill

A manual inspection of the top 100 common trees shows that the most frequent pathway (90 occurrences) that does not rely on navigational links of the infobox is a path from \textsc{Egg} to \textsc{Philosophy}, following only the first link in the article. This behavior may be caused by readers curious to verify the popular Wikipedia property that following the first link of each page recursively leads to the \textsc{philosophy} article \cite{lamprecht2016evaluating,fixed_point,ibrahim2017connecting}.



\xhdr{Frequent exit}
Complementary to the common entry points, we look at the last page that a reader loaded in the session. The most frequent article where the users leave the navigation is the article \textsc{Philosophy} which occurs as the last document more than 20K times. This article is then followed by popular content such as \textsc{Elizabeth II}, \textsc{Joe Biden}, and \textsc{2020 United States presidential election}.
To remove the popularity bias, we normalize using the total number of pageloads of the articles and limit to pages loaded at least 100 times in the month of the data collection. \textsc{Philosophy} remains a frequent last article acting as an exit point in 12.9\% of the cases, but the rank is dominated by \textsc{Ichiki Kitokurō} (31.1\%), \textsc{Are You In?: Nike+ Original Run} (27.9\%), \textsc{What's Your Favorite Color?: Remixes, B-Sides and Rarities} (25.8\%). A manual inspection of the top 100 reveals a high presence of historical figures and music albums.


\xhdr{Frequent origin}
Thanks to the referrer field of the requests that serve as roots of the trees, we can determine how readers reached Wikipedia. The majority of the rabbit hole sessions (63.8\%) start from requests coming from search engines, followed by an unspecified origin (14.1\%) that may include searches from toolbars and revisiting patterns where the user picked the article from the browser URL autocomplete. The other traffic sources are the main page (13.4\%), the Wikipedia internal search results (4.3\%), and Wikipedia in other languages than English (3.3\%). External websites --including social media-- contribute 0.5\% as the traffic source for the rabbit hole sessions.


    

\begin{figure*}[t]
\hfill
    \begin{minipage}[t]{.49\columnwidth}
        \centering
        \includegraphics[height=3.8cm]{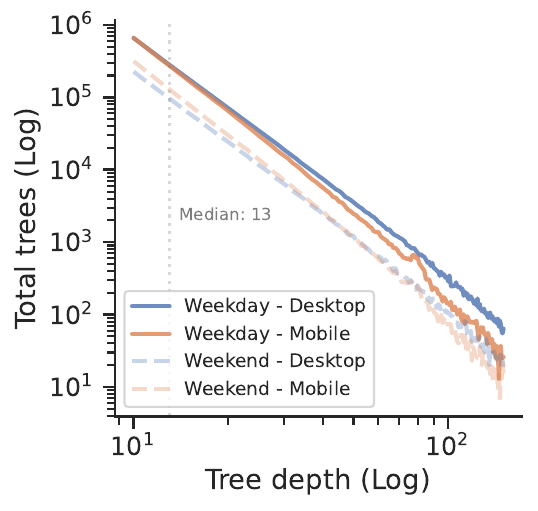}
        \subcaption{Total by depth}
        \label{fig:trees_height}
    \end{minipage}
    \hfill
    \begin{minipage}[t]{.49\columnwidth}
        \centering
        \includegraphics[height=3.8cm]{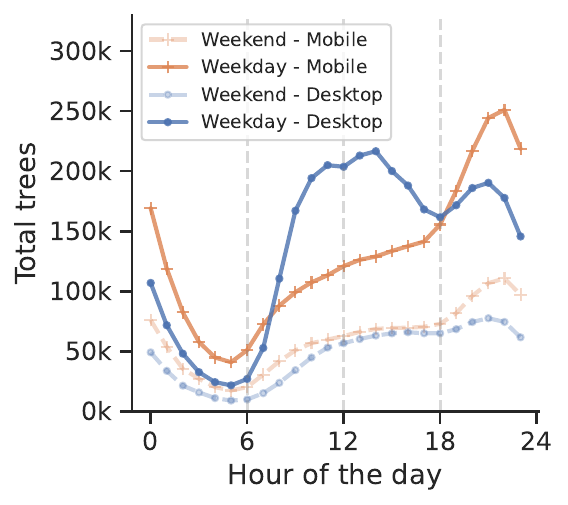}
        \subcaption{Total by time}
        \label{fig:total_rh_by_time}
    \end{minipage}
    \hfill
    \begin{minipage}[t]{.49\columnwidth}
        \centering
        \includegraphics[height=3.8cm]{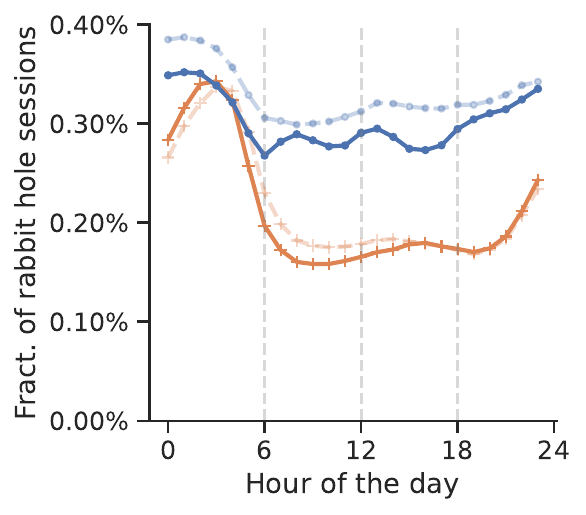}
        \subcaption{Proportion by time}
        \label{fig:proportion_by_time}
    \end{minipage}
    \hfill
    \begin{minipage}[t]{.49\columnwidth}
        \centering
        \includegraphics[height=3.8cm]{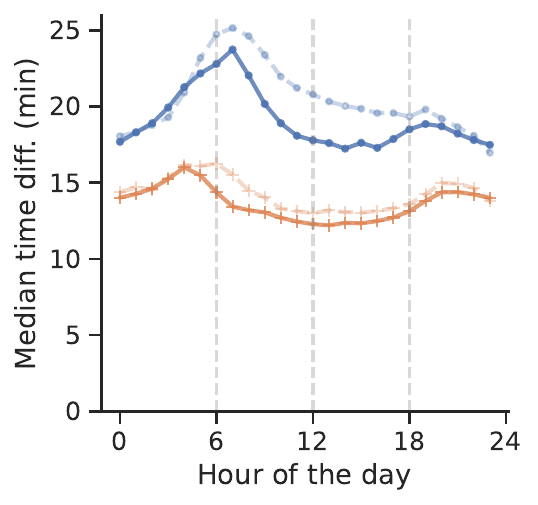}
        \subcaption{Time from first to last click}
        \label{fig:median_sessions_timediff_by_time}
    \end{minipage}
    \hfill
\caption{
(\ref{fig:trees_height}) Depth distribution of the trees with a rabbit hole pattern.  (\ref{fig:total_rh_by_time}) Total volume of wiki rabbit hole sessions by the time of the day. (\ref{fig:proportion_by_time}) Proportion of wiki rabbit hole sessions by the time of the day. (\ref{fig:median_sessions_timediff_by_time}) Median time passed from the first to the last pageload of the session. All plots show the distributions broken down by access device and weekday-weekend.
}
\label{fig:trees_stats}
\end{figure*}


\subsection{Structural patterns}
\label{subsec:structural}

As observed in previous work \cite{piccardi2021large}, the depth and size distributions of the trees generated by readers exploring Wikipedia show a long tail (approximately a straight line in a log-log plot). The trees that we retained as rabbit hole sessions have a median of 18 pageloads ($\text{Q1}=14$, $\text{Q3}=28$). We observe a small but significant\footnote{Wilcoxon signed-rank test: $p < 0.001$.\label{footnote:significant}} tendency to have larger trees from mobile ($\text{median}=19$) compared to desktop ($\text{median}=18$) and during the weekend (Saturday-Sunday -- $\text{median}=19$) compared to the working days (Monday-Friday -- $\text{median}=18$).

\xhdr{Trees depth} The median depth of the trees, i.e. the longest root-to-leaf path, is 13, meaning that half of the rabbit hole sessions do not extend beyond 12 clicks away from the first page. \Figref{fig:trees_height} shows the distribution of the tree depths. The depth of the trees does not show any significant difference based on device or day of the week.


\xhdr{Trees sizes}
The distribution of the size of navigation trees, i.e. the number of articles loaded in a session, is skewed, with an average of 24 pageloads per session and a median of 18 pageloads ($\text{Q1}=14$, $\text{Q3}=28$). By limiting to pages that acted at least 100 times as entry points of the rabbit hole, we observe that the articles that generate the largest trees are \textsc{Eurovision Song Contest 1956} ($\text{median}=42$), \textsc{Reference ranges for blood tests} ($\text{median}=38$), \textsc{List of Major League Soccer transfers 2021} ($\text{median}=36$), \textit{1st Academy Awards} ($\text{median}=35$), and \textsc{Lists of UK top-ten singles} ($\text{median}=34$).

\xhdr{Branching factor}
When readers follow multiple links on a page, they generate a fork in the navigation tree. By the trees' construction, this branching can happen when the user moves forward in the exploration by opening multiple browser tabs or backtracking the navigation to a previous point with the back button of the browser. The macro average degree of the trees is 1.36 ($\text{median}=1.2$), while trees have, on average, a maximal width considering all branches of 3.9 nodes ($\text{median}=3$).
The breakdown by device shows a higher tendency to explore multiple paths from the same article for mobile sessions ($\text{average}=1.42$, $\text{median}=1.26$) compared to desktop sessions ($\text{average}=1.30$, $\text{median}=1.15$). This observation is reflected in the average maximum width of the tree (mobile: $\text{average}=4.20$, $\text{median}=3$ vs. desktop: $\text{average}=3.58$, $\text{median}=2$).
Overall, 22\% of the trees have a chain-like structure with an average branching factor of zero.


\subsection{Temporal patterns}
\label{subsec:temporal}

\xhdr{Daily pattern}
Previous work \cite{piccardi2021large} showed that the readers follow a circadian rhythm with reduced access at night. During the working days, the requests received by English Wikipedia are balanced across desktop and mobile until the evening, when mobile access increases drastically, generating more than double the traffic of desktop devices. 

For rabbit hole sessions, we find a different pattern shown in \Figref{fig:total_rh_by_time}. The absolute number of rabbit hole sessions is significantly higher from desktop during the working daytime, whereas mobile access dominates during evening and night. 
This inversion between the desktop and mobile is not present during the weekend when mobile traffic remains the most common rabbit hole access method.

In \Figref{fig:proportion_by_time} we show the relative proportion of sessions with a rabbit hole pattern (compared to all sessions) for the two access methods throughout the day. 
In general, the fraction of rabbit hole sessions is higher at night.
On desktop devices, the portion of deep trees is consistently higher, with a further increase during the weekend. For mobile devices, the sessions started during the night (\ie 0.34\% at 3 AM) is double compared to the working hours (\ie 0.17\% at 3 PM).

\xhdr{Time spent in the rabbit hole}
We approximate the time readers spent in the rabbit hole, by computing the time difference between the first and last pageload of the navigation trees\footnote{Typically, the actual time spent reading is higher}. The median time is 15 minutes and 49 seconds ($\text{Q1}=6m42s$, $\text{Q3}=43m52s$), with differences between desktop and mobile. \Figref{fig:proportion_by_time} shows that during the day, the rabbit hole sessions started from desktop devices keep the readers consistently for more time than from mobile. Readers spend in median 5 minutes more in the rabbit hole from desktop  (18m42s vs. 13m43s)---this difference is statistically significant\footref{footnote:significant}.

% ...


\begin{figure}[t]
    \begin{minipage}[t]{.49\columnwidth}
        \centering
        \includegraphics[height=3.8cm]{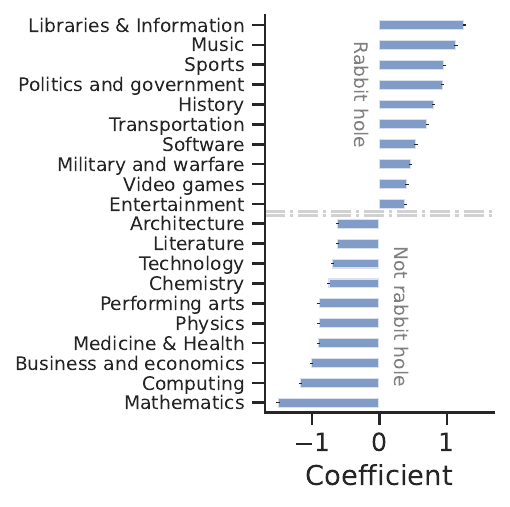}
        \subcaption{Desktop}
        \label{fig:regression_topics_desktop}
    \end{minipage}
    \hfill
    \begin{minipage}[t]{.49\columnwidth}
        \centering
        \includegraphics[height=3.8cm]{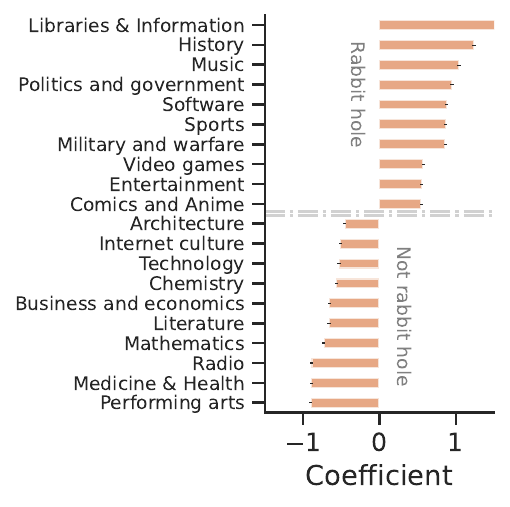}
        \subcaption{Mobile}
        \label{fig:regression_topics_mobile}
    \end{minipage}
    \hfill
\caption{
Coefficients of the logistic regressions that predict if a reader will end up in a wiki rabbit hole given the topics of the first page.
}
\label{fig:regression_coefficients}
\end{figure}

\begin{figure*}[t]
\begin{minipage}[t]{0.32\textwidth}
        \centering
        \includegraphics[height=4.2cm]{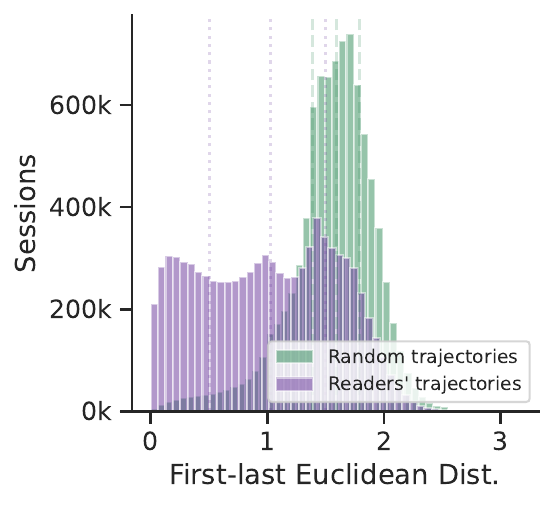}
        \subcaption{First-to-last distance}\label{fig:distances_distribution}
    \end{minipage}
    \begin{minipage}[t]{0.32\textwidth}
        \centering
        \includegraphics[height=4.5cm]{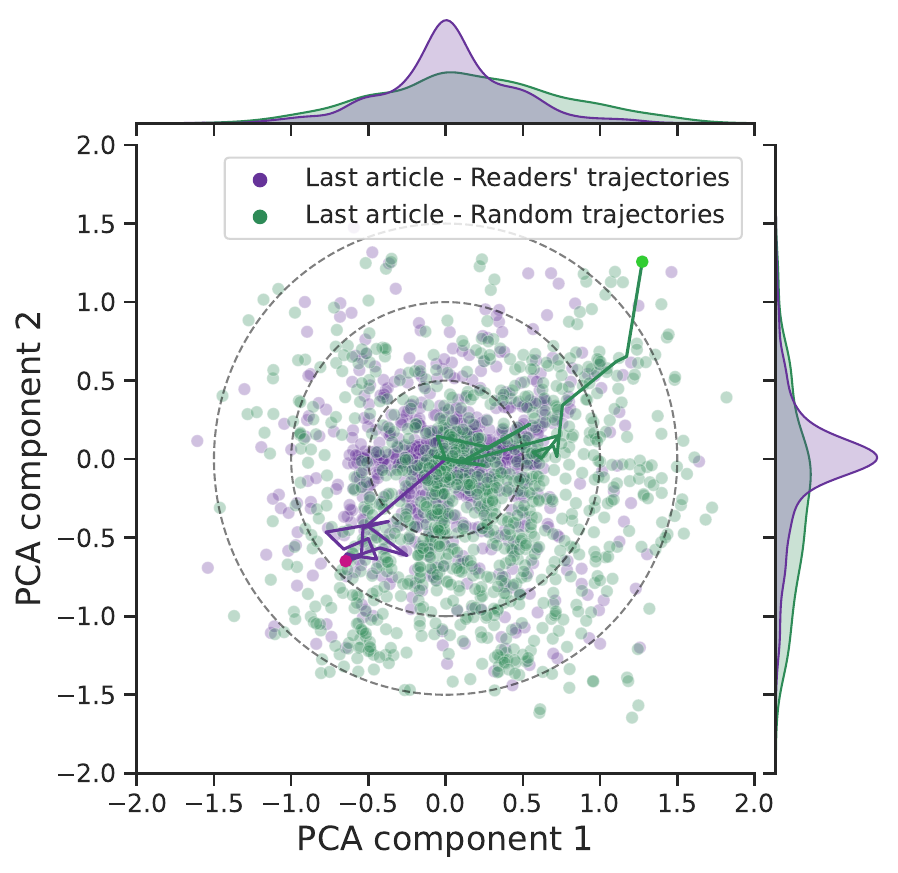}
        \subcaption{PCA projection of navigation trajectories}\label{fig:trajectories_in_ORES_space}
    \end{minipage}
    \begin{minipage}[t]{0.32\textwidth}
        \centering
        \includegraphics[height=4.2cm]{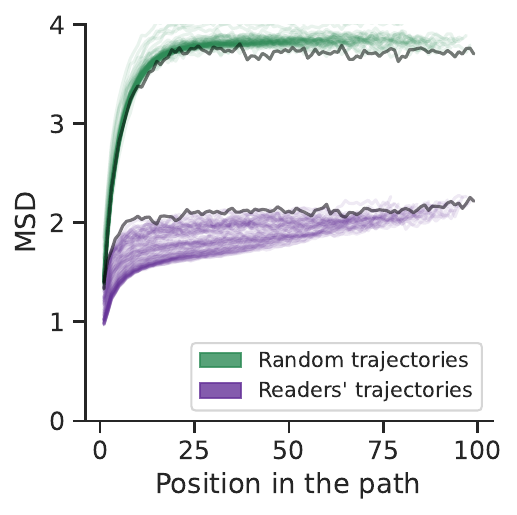}
        \subcaption{Mean squared displacement}\label{fig:topic_distance_first}
    \end{minipage}
\caption{
Sessions diffusion in topics space.  (\ref{fig:distances_distribution}): Distribution of Euclidean distances between the first and the last article of the path. Quartiles as vertical lines. (\ref{fig:trajectories_in_ORES_space}): First 2 principal components of the last article of 2000 paths generated by human readers and a random walker in the topic space defined by ORES. The first article of the session centered on the origin. Marginal plots show KDE distributions. The green and purple lines represent two examples of full trajectories.
(\ref{fig:topic_distance_first}): Mean squared displacement (MSD) of the sessions in the topic space defined by ORES. Each line represents the MSD of all the sessions of one specific length [10-100]. The dark trajectories are added for readability and represent the sessions with 100 pageviews.
}
\label{fig:topic_space}
\end{figure*}

\subsection{Topic patterns}
\label{subsec:topical}
\xhdr{Entering the rabbit hole}
To comprehend the dynamics that bring readers down the rabbit hole, we use regression analysis to study what topics are associated with the first page of the session. 
We train a logistic regression to predict if a reader will fall into the rabbit hole by using the properties (\ie topics) of the first article. We generate the dataset by selecting the first articles of all the rabbit hole sessions as positive samples and an equal number randomly picked from the non-rabbit hole trees as negative ones. This step leaves us with around 18M samples equally distributed between the two classes. Each article is then represented with a vector representing the topics probabilities obtained from ORES\footnote{\url{https://www.mediawiki.org/wiki/ORES}} \cite{ORES}. This sampling approach takes into account the popularity of the topics, ensuring that a highly popular topic is proportionally represented also in the negative class. We train two models independently for the samples generated from desktop and mobile devices, obtaining an AUC on a testing set respectively of 0.61 and 0.62.

\Figref{fig:regression_coefficients} summarises the coefficients of the topics most associated positively and negatively with the rabbit hole pattern. Similar to previous findings~\cite{piccardi2021large}, the readers who fall into a wiki rabbit hole typically start from entertainment, sport, politics, and history articles. At the same time, STEM, Medicine, and Business are overall topics where the readers engage less in deep explorations. The coefficients obtained for the regression of the two devices show variation in their ranking, but they offer qualitatively the same conclusions. 
In order to confirm the robustness of these findings, we use a linear regression model to predict the exact number of articles loaded during a navigation session, which yields qualitatively similar results (not shown).

\xhdr{Diffusion in topic space}
Often the navigation of readers falling in the rabbit hole is imagined as a long session that brings the users on a random page of Wikipedia. We verify this assumption by comparing the readers' trajectories in the topics space with a null model obtained from an unbiased random walker. We are interested in observing how the trajectories diffuse in the space with respect to the origin and if the long navigation paths converge, in multiple steps, to random clicking behavior. 
Since we need trajectories, we extract from the trees the longest path from the root to one of the leaves. If the tree has multiple longest paths of the same length, we select one random from these candidates. In 84\% of the cases, the trajectory selected is also the path leading to the last article loaded by the reader.

To compare the two sets of trajectories, we proceed in two steps: first, we create a matched dataset by running for each of the 8.9M readers-generated paths a random walk that, starting from the same article, generates a sequence of the same length. For each step, the next article is selected randomly from the list of links available on the page. Then, we assign the respective ORES topics vectors to each article visited in all trajectories.


Overall, the comparison shows that readers tend to stay semantically close to the first page, even for long sessions. \Figref{fig:trajectories_in_ORES_space} shows the PCA representation of a random sample of trajectories where the first page is centered at the origin. The marginal density distribution shows that the user-generated paths have a higher concentration close to the first page loaded when compared to the larger spread of random exploration. This intuition is reinforced by computing the Mean Squared Displacement (MSD). MSD is typically used in physics to measure the dispersion of a particle from the starting position.

MSD is formally defined\footnote{\url{https://en.wikipedia.org/wiki/Mean_squared_displacement}} as 
\begin{equation}
\label{eqn:msd}
{\rm MSD}\equiv\langle |\mathbf{x}(t)-\mathbf{x_0}|^2\rangle=\frac{1}{N}\sum_{i=1}^N |\mathbf{x^{(i)}}(t) - \mathbf{x^{(i)}}(0)|^2
\end{equation}
where $N$ is the number of trajectories, vector $\mathbf{x^{(i)}}(t)$ is the position of the navigation $i$ at step $t$, and vector $\mathbf{x^{(i)}}(0)=\mathbf{x^{(i)}_0}$ is the position of the first article for the trajectory $i$.

\Figref{fig:topic_distance_first} shows that the dispersion coefficients of the readers' navigation, stratified for different tree sizes and positions in the path, are almost half compared to the diffusion of a random walk. 

To ensure these findings are not skewed by the sessions generated from using the navigational links of the inboxes, we repeated this analysis by removing all the trees with a chain-like structure. We observe no variations in the patterns of \Figref{fig:topic_space}, concluding that readers tend to stay semantically close to the origin --compared to a random walk-- even for very long sessions.

\begin{figure}[t]
    \begin{minipage}[t]{.49\columnwidth}
        \centering
        \includegraphics[height=3.8cm]{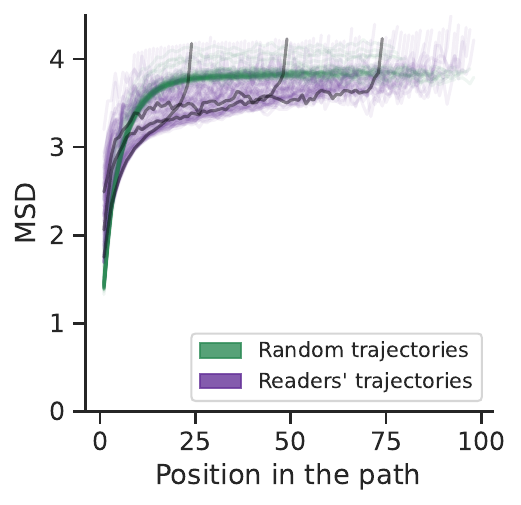}
        \subcaption{Mean squared displacement}
        \label{fig:last25_distance_from_first}
    \end{minipage}
    \hfill
    \begin{minipage}[t]{.49\columnwidth}
        \centering
        \includegraphics[height=3.8cm]{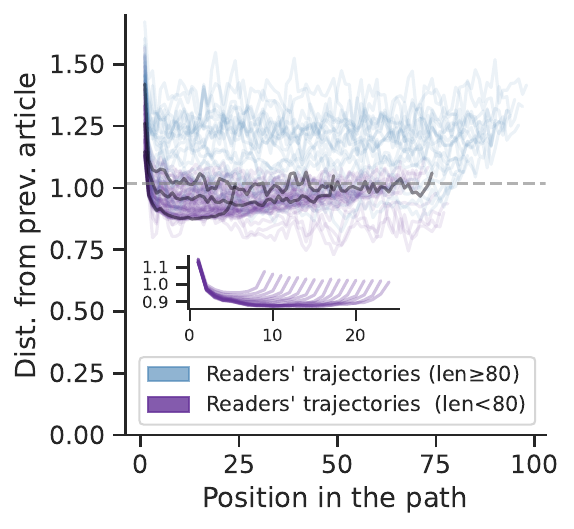}
        \subcaption{Distance from previous}
        \label{fig:last25_distance_from_prev}
    \end{minipage}
    \hfill
\caption{
(\Figref{fig:last25_distance_from_first}) Mean squared displacement of the sessions with the largest distance from the first to the last page. (\Figref{fig:last25_distance_from_prev}) Euclidean distance from the previous article in the function of the position in the path---divided into two length groups. The horizontal line represents the average distance from the previous article in the case of a random walk. Inset: zoom on the paths with a max of 25 pageloads. 
Each line represents the sessions of one specific length [10-100].
Dark lines are added for readability, and they represent the trajectories of paths of length 25, 50, and 75.
}
\label{fig:last25_stats}
\end{figure}

\xhdr{When navigation approaches random} \Figref{fig:distances_distribution} shows that a portion of the sessions that reach semantic locations comparable to a random walk on the links network. We focus on this portion of sessions by selecting the last quartile (last 25\% -- 2.2M paths) of the distribution of the Euclidean distances from the starting page. These trajectories represent the session of readers that, in the longest path, loaded an article semantically very far from the first page.

\Figref{fig:last25_distance_from_first} shows the mean square displacement of these sessions that reveals that they are close to the trajectories generated by a random walk. Interestingly, the readers' trajectories show a final steep increase in the distance from the origin, suggesting that the readers abandon the path exploration after a fast drift from the first page. This finding represents an extreme case of the behavior already observed in previous work \cite{piccardi2021large}.

MSD measures the diffusion from the original, but it does not capture the relative distance between two sequential pageloads. \Figref{fig:last25_distance_from_prev} shows how the sessions evolve. The average Euclidean distances for the consecutive pairs of articles show an initial drop indicating that readers in these special sessions, on average, jump far from the origin in the first step and then tend to move with smaller semantic jumps. Differently from the general behavior where readers favor, as next step, articles semantically close \cite{piccardi2021large,LinkSuccessfulWikipedia,StructureArticlesNavigation}, the average distance between two consecutive articles in these special sessions tends to approximate the semantic jumps of a random walk (horizontal line). Additionally, the trajectories show a difference in behavior based on the length of the path. \Figref{fig:last25_distance_from_prev} shows that very long sessions ---more than 80 clicks--- tend to have more extreme exploration patterns with jumps in semantic space that go further than a random walk. A zoom-in on the paths with less than 25 pageloads shows that the sessions tend to divergence before leaving the exploration, as previously observed in a more general study \cite{piccardi2021large}.

Finally, an investigation on time (hour and weekend) and the device used show no significant differences from the other rabbit hole sessions. 


\section{Discussion}
\label{sec:discussion}
In this paper, we have provided a first data-driven investigation on the patterns associated with long reading sessions in Wikipedia, also known as rabbit hole navigation.

\vfill

\xhdr{Summary of findings} 
By investigating the frequent paths taken by readers falling into a wiki rabbit hole, we observed that characteristics of the article layout are associated with deep navigation trees. The presence of navigational links in the infobox to transition between different instances of the same recurrent event supports the type browsing similar to reading a slideshow. 

\vfill

The dynamics of falling into a wiki rabbit hole show differences across time of the day, the device used, and the topic of the first article. The fraction of sessions with deep trees is overall higher on desktop than mobile devices and increases in both cases at night. Confirming popular belief and previous findings on more general behavior \cite{piccardi2021large}, articles about entertainment, sport, politics, and history are more common as starting points for rabbit hole sessions. 

\vfill

An investigation of the diffusion in topic space 
shows that rabbit hole sessions consist of topically coherent articles.
On average, even after long sequences of clicks, the visited article is topically much closer to the starting article than compared to a randomly chosen page.

\vfill

While most reading sessions are quite short~\cite{piccardi2021large}, here, we find that rabbit hole sessions still constitute a substantial number of sessions of readers in absolute terms -- far exceeding the number of active editors in Wikipedia. 
Overall, we show that rabbit hole sessions can exhibit distinctly different patterns in comparison to the average of all sessions which are dominated by the large number of short sessions.

\vfill

\xhdr{Limitations and future work} 
This study can spark future work to comprehend the knowledge consumption patterns and to inform the organization of Wikipedia's content.
One important limitation of the present work is that defining the rule to identify a rabbit hole session is not a trivial task. The rabbit hole concept is mainly based on anecdotal examples, and what constitutes a rabbit hole session may be subjective. We focus on the trees with at least a depth of 10 nodes, but other approaches, for example, could employ methods based on the divergence of the reader from the origin or on time spent reading articles, regardless of the number of steps. 
Another future aspect worth exploring is the diversity of behavior based on geographical and cultural factors. Our analysis is limited to the English edition of Wikipedia, but we are interested in extending it to multiple languages. Additionally, further interactions with the page, such as previews and dwelling time, can be taken into account to paint a more complete picture of the readers' navigation in the rabbit hole. Finally, these findings can be used to better serve readers' needs. For example, the chain-like navigation using the links in the infobox could suggest either i) a desire of the readers to have a more complete overview on a set of articles, or ii) difficulty in finding the article containing relevant content through traditional text-based search. 

\xhdr{Conclusion} 
Readers visiting Wikipedia cover a wide spectrum in terms of their motivations, needs, and prior knowledge~\cite{lemmerich2019why}.
In this study, we focused on a specific setting in which readers embark on in-depth navigation in Wikipedia, i.e. rabbit hole sessions. 
The characteristics of these sessions differ from the majority of very short sessions suggesting that rabbit hole sessions satisfy succinctly distinct needs of readers.
This work thus provides new quantitative insights into how Wikipedia is used by readers which could empower the community to make informed decisions around the organization of Wikipedia's content.
More generally, we hope to inspire future research on online knowledge consumption and add a small piece to our understanding of Wikipedia readership. 


\begin{acks}
This project was partly funded by the Swiss National Science Foundation (grant 200021\_185043), the European Union (TAILOR, grant 952215), and the Microsoft Swiss Joint Research Center. We also gratefully acknowledge generous gifts from Facebook and Google supporting West's lab.
\end{acks}

\balance
\bibliographystyle{ACM-Reference-Format}
\bibliography{references}

\end{document}
\endinput

%% file: dlab.tex
\usepackage{hyphenat}
\usepackage{xspace}
\usepackage{amsmath}
\usepackage{amsfonts}
\usepackage{url}
\usepackage{marvosym}
\usepackage{ifthen}
\theoremstyle{plain}

\newcommand{\chatoDisplayMode}[1]{#1}

\definecolor{MyRed}{rgb}{0.6,0.0,0.0} 
\definecolor{MyBlack}{rgb}{0.1,0.1,0.1} 
\newcommand{\inred}[1]{{\color{MyRed}\sf\textbf{\textsc{#1}}}}
\newcommand{\frameit}[2]{
  \begin{center}
  {\color{MyRed}
  \framebox[.9\columnwidth][l]{
    \begin{minipage}{.85\columnwidth}
    \inred{#1}: {\sf\color{MyBlack}#2}
    \end{minipage}
  }\\
  }
  \end{center}
}

\newcommand{\note}[2][]{\chatoDisplayMode{\def\@tmpsig{#1}\frameit{{\Pointinghand} Note}{#2\ifx \@tmpsig \@empty \else \mbox{ --\em #1}\fi}}}
\newcommand{\todo}[2][]{\chatoDisplayMode{\def\@tmpsig{#1}\frameit{{\Writinghand} To-do}{#2\ifx \@tmpsig \@empty \else \mbox{ --\em #1}\fi}}}

\newcommand{\abbrevStyle}[1]{#1}

\newcommand{\ie}{\abbrevStyle{i.e.}\xspace}

\newcommand{\Secref}[1]{Sec.~\ref{#1}}

\newcommand{\Figref}[1]{Figure~\ref{#1}}

\newcommand{\xhdr}[1]{\vspace{1.7mm}\noindent{{\bf #1.}}}

\newcommand{\textcite}[1]{\citeauthor{#1} \shortcite{#1}}

\newcommand{\hide}[1]{}

\makeatletter
\newcommand{\iffont}[2]{\ifthenelse{\equal{\f@family}{#1}}{#2}{}}
\makeatother

\iffont{ptm}{
  \usepackage{mathptmx}

  \DeclareSymbolFont{greek}{OML}{cmm}{m}{n}
  \DeclareMathSymbol{\alpha}{\mathalpha}{greek}{"0B}
  \DeclareMathSymbol{\beta}{\mathalpha}{greek}{"0C}
  \DeclareMathSymbol{\gamma}{\mathalpha}{greek}{"0D}
  \DeclareMathSymbol{\delta}{\mathalpha}{greek}{"0E}
  \DeclareMathSymbol{\epsilon}{\mathalpha}{greek}{"0F}
  \DeclareMathSymbol{\zeta}{\mathalpha}{greek}{"10}
  \DeclareMathSymbol{\eta}{\mathalpha}{greek}{"11}
  \DeclareMathSymbol{\theta}{\mathalpha}{greek}{"12}
  \DeclareMathSymbol{\iota}{\mathalpha}{greek}{"13}
  \DeclareMathSymbol{\kappa}{\mathalpha}{greek}{"14}
  \DeclareMathSymbol{\lambda}{\mathalpha}{greek}{"15}
  \DeclareMathSymbol{\mu}{\mathalpha}{greek}{"16}
  \DeclareMathSymbol{\nu}{\mathalpha}{greek}{"17}
  \DeclareMathSymbol{\xi}{\mathalpha}{greek}{"18}
  \DeclareMathSymbol{\pi}{\mathalpha}{greek}{"19}
  \DeclareMathSymbol{\rho}{\mathalpha}{greek}{"1A}
  \DeclareMathSymbol{\sigma}{\mathalpha}{greek}{"1B}
  \DeclareMathSymbol{\tau}{\mathalpha}{greek}{"1C}
  \DeclareMathSymbol{\upsilon}{\mathalpha}{greek}{"1D}
  \DeclareMathSymbol{\phi}{\mathalpha}{greek}{"1E}
  \DeclareMathSymbol{\chi}{\mathalpha}{greek}{"1F}
  \DeclareMathSymbol{\psi}{\mathalpha}{greek}{"20}
  \DeclareMathSymbol{\omega}{\mathalpha}{greek}{"21}
  \DeclareMathSymbol{\varepsilon}{\mathalpha}{greek}{"22}
  \DeclareMathSymbol{\vartheta}{\mathalpha}{greek}{"23}
  \DeclareMathSymbol{\varpi}{\mathalpha}{greek}{"24}
  \DeclareMathSymbol{\varrho}{\mathalpha}{greek}{"25}
  \DeclareMathSymbol{\varsigma}{\mathalpha}{greek}{"26}
  \DeclareMathSymbol{\varphi}{\mathalpha}{greek}{"27}
  \DeclareSymbolFont{otone}{OT1}{cmr}{m}{n}
  \DeclareMathSymbol{\Gamma}{\mathalpha}{otone}{0}
  \DeclareMathSymbol{\Delta}{\mathalpha}{otone}{1}
  \DeclareMathSymbol{\Theta}{\mathalpha}{otone}{2}
  \DeclareMathSymbol{\Lambda}{\mathalpha}{otone}{3}
  \DeclareMathSymbol{\Xi}{\mathalpha}{otone}{4}
  \DeclareMathSymbol{\Pi}{\mathalpha}{otone}{5}
  \DeclareMathSymbol{\Sigma}{\mathalpha}{otone}{6}
  \DeclareMathSymbol{\Upsilon}{\mathalpha}{otone}{7}
  \DeclareMathSymbol{\Phi}{\mathalpha}{otone}{8}
  \DeclareMathSymbol{\Psi}{\mathalpha}{otone}{9}
  \DeclareMathSymbol{\Omega}{\mathalpha}{otone}{10}
  \DeclareSymbolFont{syms}{OML}{cmm}{m}{it}
  \DeclareMathSymbol{\partial}{\mathord}{syms}{"40}
  \DeclareMathAlphabet{\mathbold}{OML}{cmm}{b}{it}
  \DeclareSymbolFont{largesymbols}{OMX}{cmex}{m}{n}

}